\documentstyle[11pt,newpasp,twoside,epsfig]{article}
\begin{document}

\title{X-ray Emission from Isolated Be Stars}

\author{David H. Cohen}
\affil{Bartol Research Institute, University of Delaware, Newark, DE 19716}

\begin{abstract}

I discuss the X-ray observations of Be stars, and how their properties compare 
to non-emission B stars.  I focus on several specific stars that show high flux 
levels and variability but also report on several interesting survey results.  
The Be X-ray properties are discussed in the context of wind-shock X-ray 
emission from normal OB stars as well as in the context of general mechanisms 
that have been proposed to explain the Be phenomenon.  Finally, I conclude with 
a discussion of the spectral diagnostics that will be available from the new 
generation of X-ray telescopes.

\end{abstract}


\section{Introduction}

The X-ray emission from single Be stars must be viewed in the context of OB star 
X-ray emission.  The X-rays that are ubiquitous in O stars at a level of $L_x 
\sim 10^{-7}L_{Bol}$ are thought to come from shocks in the massive winds of 
these stars.  The generally accepted shock mechanism is the line-force 
instability (Lucy \& Solomon 1970; Owocki, Castor, \& Rybicki 1988), which 
arises directly from the physics of line driving and leads to an abundance of spatial structure, even without any external perturbations.  

In this paper I will explore the extent to which the X-ray properties of Be 
stars can understood in terms of the wind-shock emission that is invoked to 
explain the X-rays emitted by O and early B stars.  
I will contrast the observed X-ray emission from Be stars to that seen in non-
emission B stars and focus on several specific stars.  This paper will explore 
what constraints these observations can put on models of Be activity and also 
look forward to the new diagnostics that will be available with the next 
generation of X-ray telescopes. 

\section{X-rays from Hot Stars}

The wind-shock paradigm has not been observationally tested in any great detail, 
as the only X-ray data available have been of very limited spectral resolution.  
Some general, phenomenological properties of OB star X-ray emission have been 
determined via both large surveys (primarily with {\it ROSAT}) and detailed 
measurements of a few stars with the higher resolution instruments {\it ASCA} 
and {\it EUVE}.   In Table \ref{Tab:missions} I summarize the important 
properties of EUV and X-ray telescopes past, present, and future. 

\begin{table}
\caption{X-ray and EUV Spectral Missions.} \label{Tab:missions}
\begin{center}
\begin{tabular}{ccccc}
\tableline
\tableline
Mission & Launch & Spectral Range & Resolution & Effective Area \\
 & & (keV) & ($\lambda / \Delta\lambda$) & (cm$^2$) \\
\tableline
{\it Einstein} & 1978 & 0.4 - 4.0 & 2 & 100 \\
{\it ROSAT PSPC} & 1990 & 0.1 - 2.3 & 2 & 200 \\
{\it EUVE} & 1992 & 0.02 - 0.18 & 300 & 1 \\
{\it ASCA} & 1993 & 0.5 - 12 & 20 & 1000 \\
{\it RXTE PCA} & 1995 & 2 - 60 & 10 & 6500 \\
{\it Chandra} gratings & 1999 & 0.1 - 10 & 600 & 20 \\
{\it XMM RGS} & 2000 & 0.4 - 2 & 300 & 100 \\
\tableline
\end{tabular}
\end{center}

\end{table}

In several hot stars, X-ray emission lines have been measured, confirming the 
thermal nature of the emission.  Measurements of the relative lack of 
attenuation of the soft\footnote{In this paper, the term ``soft X-ray'' refers 
to X-rays with $h\nu < 1$ keV, roughly, while ``hard X-ray'' refers to $h\nu > 
1$ keV.} X-rays in several O stars (Cassinelli et al. 1981; Cassinelli \& Swank 
1983; Hillier et al. 1993) and the EUV in the B giant $\epsilon$ CMa (Cohen et 
al. 1996), as well as modeling of the O VI line in $\zeta$ Pup (MacFarlane et 
al. 1993) has shown that the X-ray emitting plasma in these stars must be 
spatially distributed throughout the stellar wind. Analyses of the spectral 
energy distributions in the low-resolution {\it Einstein} (Hillier et al. 1993) 
and {\it ROSAT} data (Cohen, Cassinelli, \& MacFarlane 1997), and of specific 
emission line ratios in the higher-resolution {\it EUVE} data (Cohen et al. 
1996), showed that in most hot stars the temperature distribution of X-ray 
emiting plasma is weighted toward less-hot material.  The most detailed 
observations show that the vast majority of plasma has $T < 10^6$ K, and there 
is relatively little with $T > 5 \times 10^6$ K (Cohen et al. 1996). 

The overall levels of X-ray emission imply that a small fraction of the wind 
mass is heated to high temperatures at any given time in O stars (Long \& White 
1980; Hillier et al. 1993; Owocki \& Cohen 1999).  However, for B stars, this 
filling factor of hot gas is larger, and for mid- and late-B stars it can exceed 
unity, assuming the theoretical mass-loss rates are correct (Berghofer \& 
Schmitt 1994; Cohen et al. 1997). This presents a serious challenge to most 
wind-based models of X-ray production in these B stars. Of course, clumping of 
the wind can, in theory, alleviate this discrepancy because thermal X-ray 
emission scales as density-squared.  However, prodigious amounts of clumping 
would be required in many cases, and keeping overdense material in the wind very 
hot is a difficult task.

Finally, there has been very little time-variability observed in the X-ray 
emission of early-type stars, especially when compared to observations of late-
type stars.  This observational fact implies that the emitting structures, 
whatever their origin, are either in a steady-state configuration, or are so 
numerous that their aggregate properties are constant. 

This observational database is generally in qualitative agreement with the 
predictions of the line-force instability wind shocks in terms of temperature 
and spatial structure, at least for O and very early B stars.  However, other 
models might also fit the limited data.  There have, in fact, been quite a few 
models proposed for the production of X-rays in OB stars.  These include 
alternate wind-shock models (Mullan 1984; MacFarlane \& Cassinelli 1989; Porter 
\& Drew 1995; Cranmer \& Owocki 1996), magnetic/coronal models (Waldron 1984; 
Tout \& Pringle 1995), and magnetic-wind hybrid models (Babel \& Montmerle 
1997a).  Magnetic models are attractive because they are theoretically capable 
of producing the large observed emission measures.  Some of these alternative 
scenarios might be especially applicable to Be stars. 

There are several reasons why we might expect Be stars to have different X-ray 
properties than normal OB stars. The first is that Be stars tend to have denser 
and more variable winds than non-emission B stars. Because X-ray emission is 
proportional to density squared, a modest increase in mass-loss rate or a 
deviation from a smooth, spherically symmetric flow would lead to an increase 
the X-ray luminosity.  The second reason that Be stars might be expected to 
have enhanced X-ray emission is the occurrence of non-radial pulsations (NRPs) 
in many of these objects.  Such photospheric variability might lead to the 
formation of co-rotating interaction regions and associated shocks and X-ray 
emission. NRPs could also make the formation of clumps in the wind more likely.

If the Be phenomenon involves magnetic fields, these could provide an additional
possible source of X-ray activity.  There are currently no viable detailed 
models of coronal-type activity in Be stars, but if magnetic fields are present 
in some Be stars, this would at least suggest the possibility of magnetic 
heating and/or confinement.  One model that has been applied to the purported 
magnetic rotator $\theta^1$ Ori C is a magnetically confined wind, in which a 
radiation driven outflow is trapped in a closed magnetic loop, within which the 
flows that originate at the two foot-points are forced to collide at the top of 
the loop with a large relative velocity (Babel \& Montmerle 1997a,b).  It has 
recently been shown that at least one Be star, $\beta$ Cep, has a large-scale 
magnetic field with a polarity that varies with the stellar rotation (Henrichs 
et al. 2000).  

Finally, the presence of the Be disks themselves could be related to X-ray 
production, possibly in combination with any of the other general mechanisms 
already discussed, and perhaps intimately connected with the cause of the 
episodic mass-loss that appears to feed the Be disks.  In any scenario in which 
the stellar wind feeds the disk, such as the wind compressed disk (WCD) model 
(Bjorkman \& Cassinelli 1993), there is potential for X-ray emission at the 
wind-disk interface.

\section{X-ray Data for Be Stars}

\subsection{Surveys}

The {\it ROSAT All-sky Survey (RASS)} was used to search for white dwarf 
companions to Be stars, and as a byproduct compiled information about single Be 
stars (Meurs et al. 1992). The results of the survey indicated that the 
incidence of X-ray emission is lower in Be stars than in B stars (6 \% vs. 16 \%).  However, because 
the {\it RASS} involved mostly very brief pointings, it detected only the 
brightest X-ray sources among hot stars (typical limiting luminosity of $L_X 
\approx 10^{30}$ ergs s$^{-1}$).  

\begin{figure}
\plotfiddle{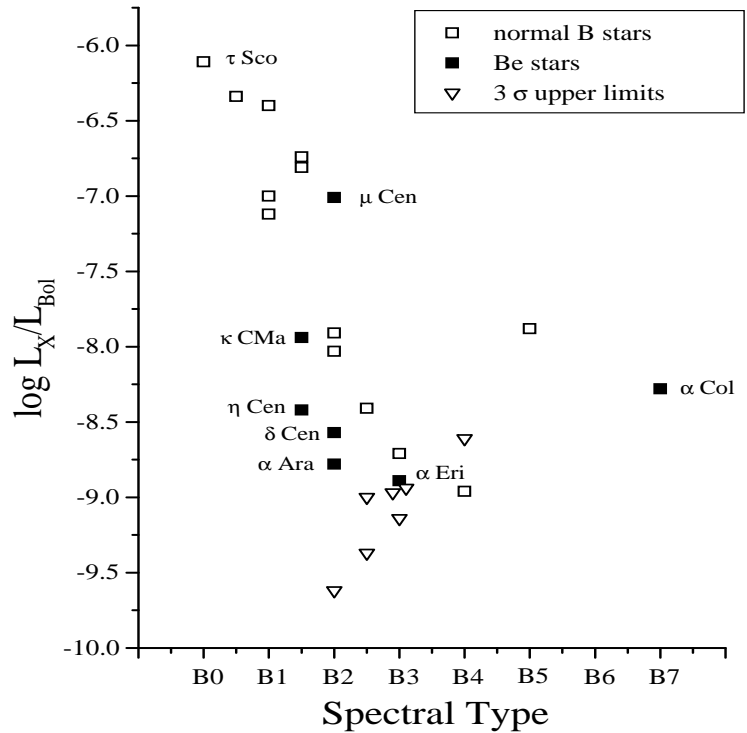}{3.75in}{0}{55}{45}{-150}{-30}
\caption{The trend of $L_X/L_{Bol}$ with spectral subtype among B Stars.} 
\label{Fig:lxlb}
\end{figure}

\begin{figure}
\plotfiddle{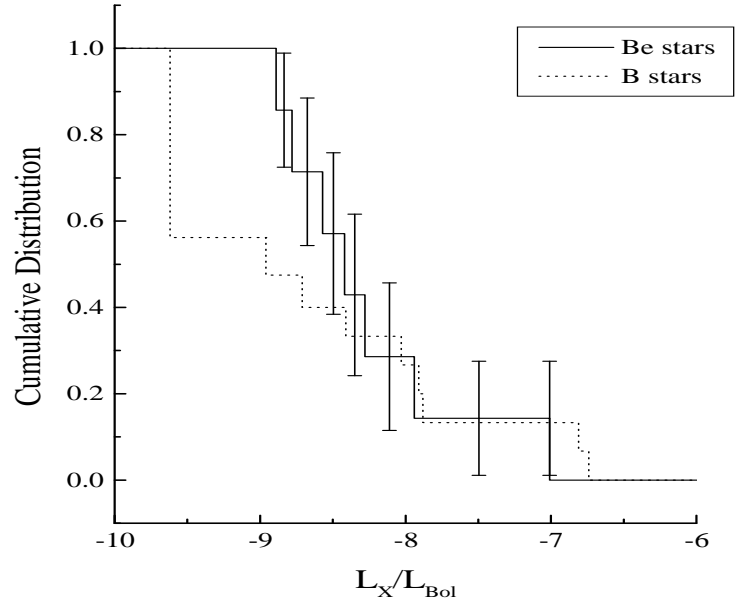}{3.00in}{0}{50}{36}{-135}{-27}
\caption{Cumulative $L_X/L_{Bol}$ distributions of Be and B stars reconstructed 
from the pointed {\it ROSAT} observations using a Kaplan-Meier estimator 
implemented in the ASURV survival analysis package (Feigelson \& Nelson 1985).} 
\label{Fig:cumulative}
\end{figure}

A much more sensitive survey of a smaller number of B stars was carried out 
using pointed {\it ROSAT PSPC} observations (Cohen et al. 1997).  The limiting 
sensitivity in this survey was better than $L_X \approx 10^{28}$ ergs s$^{-1}$.
Unlike the Meurs et al. (1992) study, in which any star having had $H{\alpha}$ 
in emission was called a Be star, the Cohen et al. (1997) determined the 
current Be status from recent literature.  Cohen et al. detected all 7 of 
the Be stars but only 8 out of 15 non-emission Be stars of the 
same spectral types.  In Figure \ref{Fig:lxlb} I show the 
$L_X/L_{Bol}$ values the survey stars as a function of spectral type.  
In Figure \ref{Fig:cumulative} the cumulative $L_X/L_{Bol}$ distributions of the 
Be and B stars are compared.  
The median value for Be stars is about three times higher than for the B stars.  


\subsection{Individual Stars}

There are several interesting individual cases of Be star X-ray emission which 
I will discuss in this subsection.  First, it should be mentioned that the 
unusual Be star $\gamma$ Cas will not be discussed in detail.  
Its X-ray properties are so extreme as to not be applicable to the understanding
of many other Be stars.  As such, it probably deserves an entire article to 
itself, and, in any case, many of its properties are discussed by other authors 
at this conference (e.g. Smith \& Robinson, these proc.).

In general, OB stars show remarkably constant 
X-ray emission, with only a handful of detections of time variability.  As 
mentioned above, this is surprising in the context of
wind shocks, as the growth and decay of individual shocks might be expected to 
lead to significant variability.  A significant fraction of the small number of X-ray variable hot stars, however, are Be stars.  

\begin{figure}
\plotfiddle{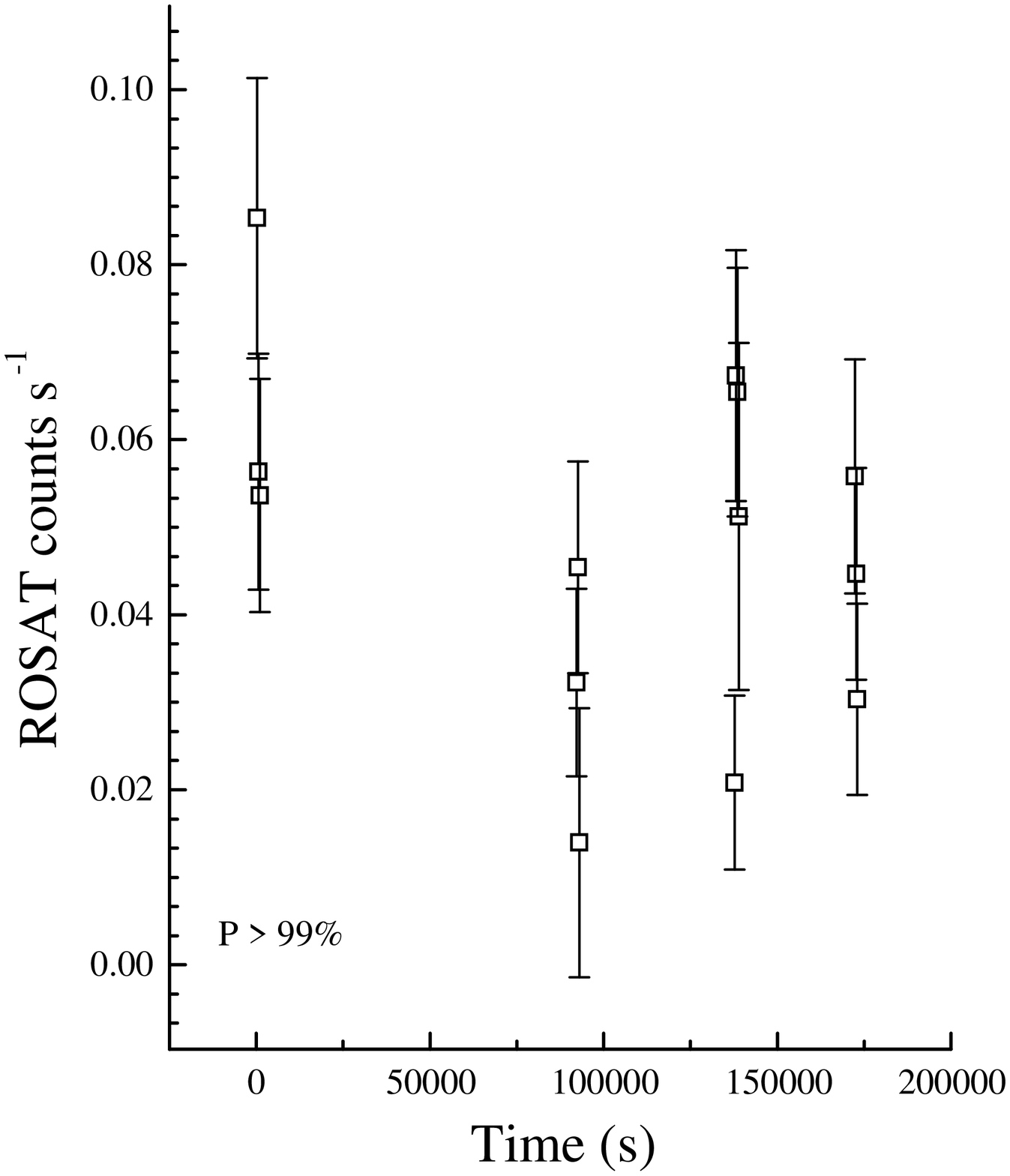}{6.25in}{0}{55}{55}{-180}{-50}
\plotfiddle{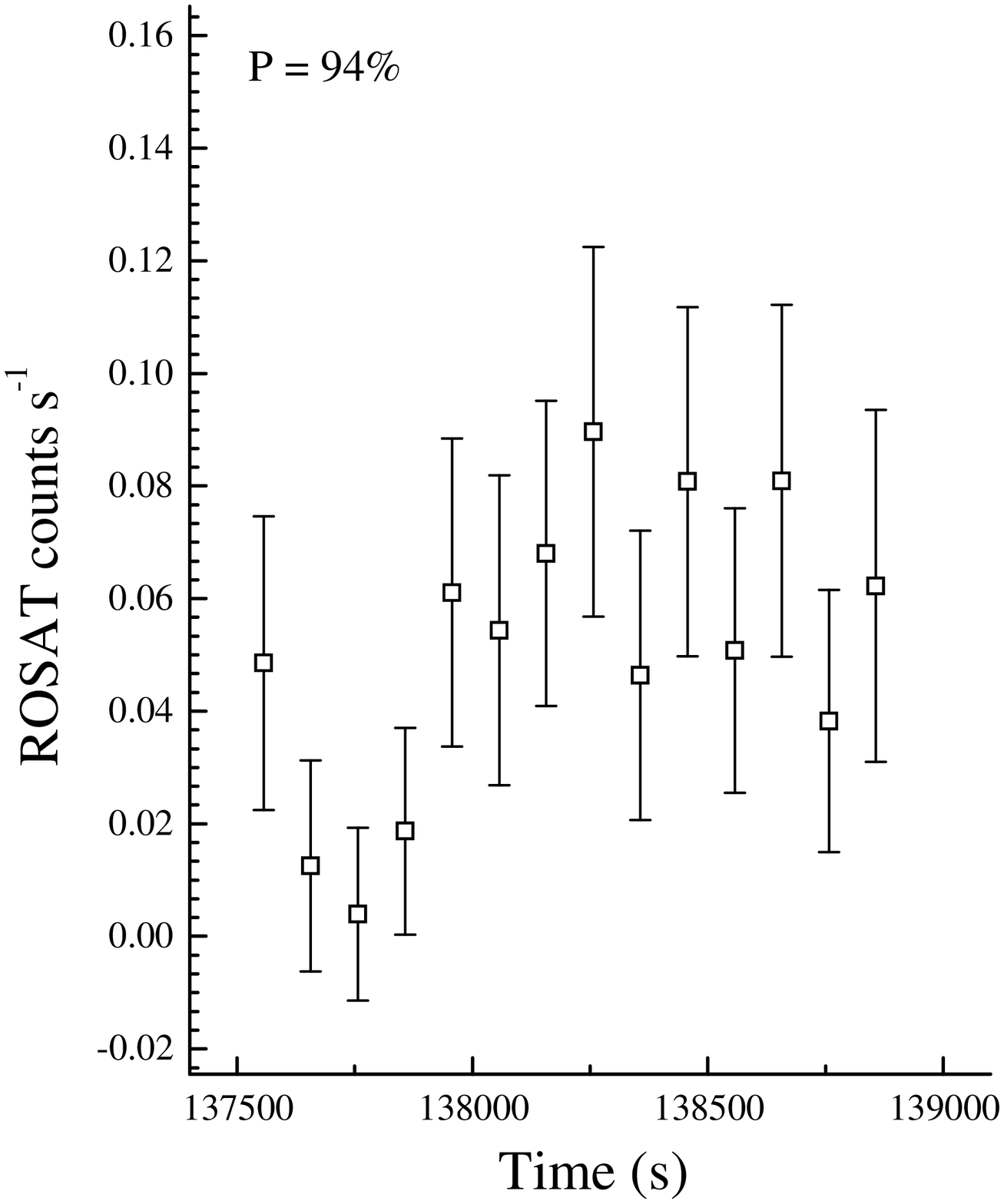}{0in}{0}{32}{32}{-20}{245}
\caption{{\it ROSAT PSPC} count rate for $\mu$ Cen.  Each data point represents 
about 400 s.  The hypothesis of a constant source can be rejected at better than 
the 99 \% level for the observation as a whole.  The data within each individual 
orbit, however, is relatively constant, with the most significant variability 
seen in the data from the third orbit (inset), for which the constant-source 
hypothesis can be rejected at the 94 \% level.} \label{Fig:mcen}
\end{figure}

The B2 Ve star $\mu$ Cen is one of only two B stars out of the 27 in the {\it 
ROSAT} sample of Cohen et al. (1997) to show time variability.  In Figure 
\ref{Fig:mcen} observations are shown that were taken during four orbits, 
spanning about two days.  
There is a strong indication of variability on timescales of more than 1000 s, 
but little firm evidence for variability on shorter timescales.  These data were 
taken during 1993, when the optical emission activity of the star was on the 
increase.  However, this {\it ROSAT} observation was {\it not} made during an 
outburst, according to the NRP ephemeris (Rivinius et al. 1998). 

The higher X-ray flux observed at the beginning of the {\it ROSAT} observation 
of $\mu$ Cen was due almost entirely to an excess in the hard band ($h{\nu} > 
0.5$ keV).  In fact, the spectral energy distribution of this star over the 
entire duration of the observation is relatively hard.  Only $\tau$ Sco and 
$\xi^1$ CMa have harder {\it ROSAT} spectra among the 27 B stars observed by 
Cohen et al. (1997). 

\begin{figure}
\plotfiddle{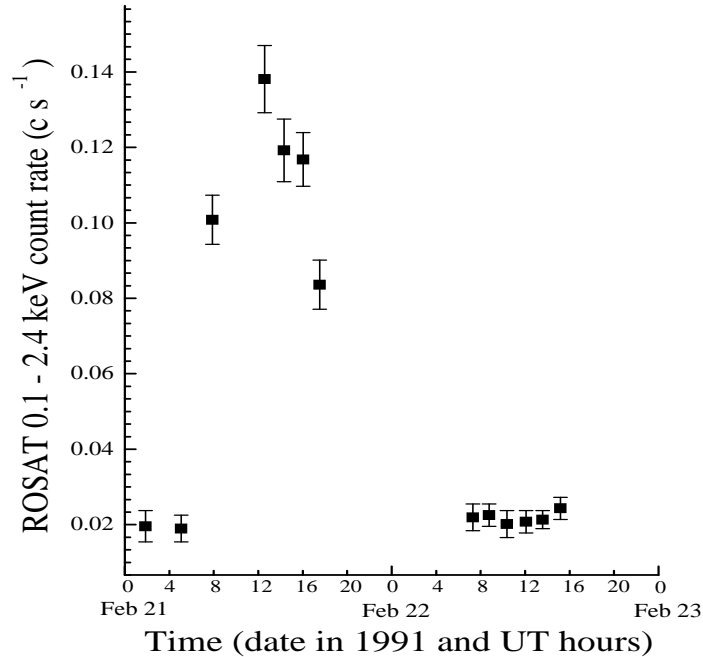}{3.35in}{0}{50}{40}{-135}{-30}
\caption{{\it ROSAT PSPC} X-ray light curve for $\lambda$ Eri (from Smith et al. 
1993, used with permission).} 
\label{Fig:leri}
\end{figure}

The most significant X-ray variability observed in any hot star was the large X-ray 
flare in $\lambda$ Eri (B2 IIIe), which lasted slightly less than 1 day, and
represented a seven-fold increase in the {\it ROSAT PSPC} count rate (Smith et 
al. 1993).  The total fluence in this event was in excess of $10^{36}$ ergs, 
making it the largest stellar X-ray flare ever detected (see Figure 
\ref{Fig:leri}).
As in $\mu$ Cen, the increase in the X-ray flux is attributable almost entirely 
to the hard {\it ROSAT} band.  Smith et al. (1997) have also found evidence for 
dense, temporary structures above the photosphere of this star.  Combined with 
the hard spectral response of the X-ray event, these data provide a picture of a 
Be star with surface magnetic fields and associated magnetic heating.  

\begin{figure}
\plotfiddle{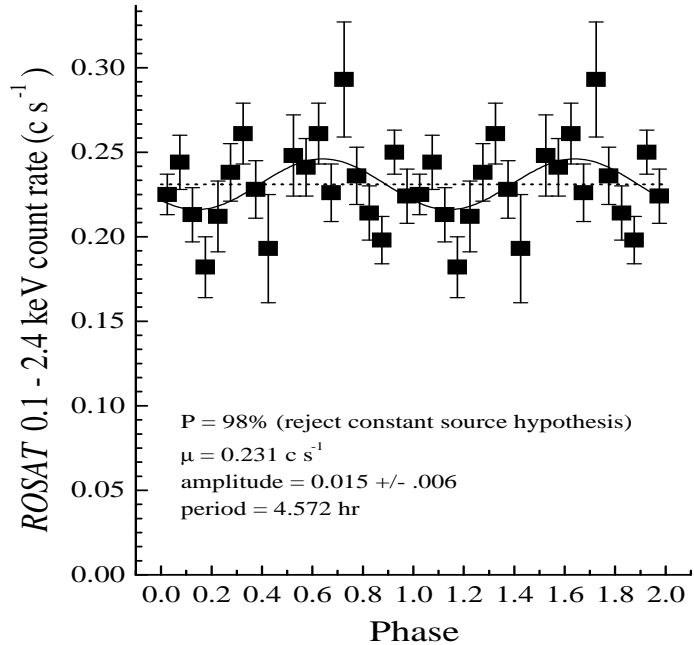}{3.25in}{0}{52}{40}{-146}{-38}
\caption{{\it ROSAT PSPC} X-ray light curve for $\beta$ Cep, folded on the 4.572 
hour primary optical period.  The hypothesis of a constant source can be 
rejected at the 98 \% confidence level.} 
\label{Fig:bcep}
\end{figure}

The prototype $\beta$ Cephei variable, $\beta$ Cep (B1 IIIe), which also shows 
H$\alpha$ emission, is another Be star that has X-ray variability.  The level 
of variability is low, but it is periodic, and is modulated on the same period 
as the optical variability associated with its NRPs (Cohen, Finley, \& 
Cassinelli 2000). This relationship between NRPs and photospheric variability on 
the one hand, and presumably wind-related X-ray activity on the other, is one of 
the strongest indications yet of a direct photosphere-wind connection in hot 
stars.  Furthermore, it is evidence that pulsations in hot stars can affect the 
circumstellar environment and the X-ray emission in these objects.  The phased 
X-ray lightcurve is shown in Figure \ref{Fig:bcep}. 

The variability of the X-ray emission of $\beta$ Cep may not be directly 
associated with its status as a Be star, as Cohen et al. (2000) find evidence 
for similar variability in 3 other $\beta$ Cephei variables, none of which are 
Be stars.  It should be noted that the time coverage of the {\it ROSAT} 
observation of this star is exceptionally good, and it is certainly possible 
that more OB stars would have modest X-ray variability detected if they could be 
observed for 20,000 s, as $\beta$ Cep was.  It should also be noted that while 
this observation was long, it was not long enough to detect modulation on the 12
day period over which the magnetic field, recently detected by Henrichs et al. 
(2000), has been found to vary. 

Finally, I will briefly mention $\tau$ Sco (B0 V), which has been suggested to 
be a Be star with a weak disk seen pole-on (Waters et al. 1993).  However, it 
now seems that the observed Brackett emission lines can be better explained by a 
non-LTE effect at the base of the wind, rather than by a disk (Waters, these 
proc.).  The X-ray properties of this star are relevant, however, because 
they show that certain hot stars have X-ray emission that simply cannot be 
explained by the standard line-force instability wind shock mechanism.  The 
X-ray emission from $\tau$ Sco is so hard that a significant quantity of plasma 
with temperatures in excess of 20 million K must be present on this star (Cohen, 
Cassinelli, \& Waldron 1997).  This might be due to magnetic activity or 
magnetically confined winds.  The presence of a magnetic field on this hot star 
is plausible due to its extreme youth (Kilian 1992).  Alternatively, the hard X-rays 
might be produced by an interaction between dense, infalling blobs (seen in 
red-shifted O {\small VI} absorption), and the fast stellar wind 
(Howk et al. 2000).  


\section{Constraints on Be Star Models Provided by the X-ray Data}

Based on the available X-ray database of OB stars, it can be concluded that the 
X-ray properties of Be stars are not very different than those of B stars in 
general.  Their X-ray luminosities may be somewhat higher (Cohen et al. 1997 
suggest a factor of three, but their sample may not be completely unbiased).  
And there are several interesting cases of stronger, harder, and more 
time-variable X-rays from Be stars.  However, it should be kept in mind that the 
existing data are relatively sparse, with only about 25 isolated Be stars 
detected with X-ray telescopes over the past two decades, and with very few of 
these observations having durations in excess of 5000 s. 

To the extent that Be stars are moderately more X-ray-active than non-emission B 
stars, there are several possible causes.  The higher luminosities could be due 
to the higher mass-loss rates that Be stars have compared to B stars of the same 
spectral subtype (Grady, Bjorkman, \& Snow 1987).  Alternately, the Be X-rays 
could be due to wind shocks that are enhanced compared to those in B star winds, 
perhaps by additional clumping, driving at the base via NRPs, or by interactions 
with corotating interaction regions or with the Be disks.  

The time-variable X-ray fluxes and high X-ray temperatures seen in $\mu$ Cen and 
$\lambda$ Eri, among other stars, have several possible implications. The time 
variability may be indicative of magnetic flaring (more likely in the case of 
$\lambda$ Eri, in which the X-ray outburst was large) or of emission from an 
ensemble of wind shocks that is dominated by a small number of shock zones.  The 
high X-ray temperatures (in $\tau$ Sco, but also in the X-ray enhancements seen 
in $\lambda$ Eri and $\mu$ Cen) may also be indicative of magnetic activity, or 
hybrid magnetic-wind activity.  Alternately, they could be due to shocks 
involving the winds of these Be stars interacting with slow, dense (or even 
infalling) clumps or with the disk. 

The presence of disks around Be stars might be expected to affect the X-ray 
properties of these objects via interactions with the stellar wind.  In the WCD 
model, or any other model in which wind material directly feeds a disk, the wind 
is decelerated across a shock as it enters the disk.  Specific models of the 
wind compressed disk predict shock temperatures of several hundred thousand 
degrees Kelvin, with somewhat higher temperatures in stars rotating at close to the break-up velocity (Bjorkman \& Cassinelli 1993).  If Be disks have an inner 
radius ({\it i.e.} they are detached from the stellar surface) then there may 
also be shock emission where the wind impacts the inner edge of the disk.  
Additionally, there may be infall of material from a disk that cannot be 
rotationally supported, and this material may interact either with the wind or 
the photosphere itself, leading to shocked plasma and X-ray emission.  

\begin{figure}
\plotfiddle{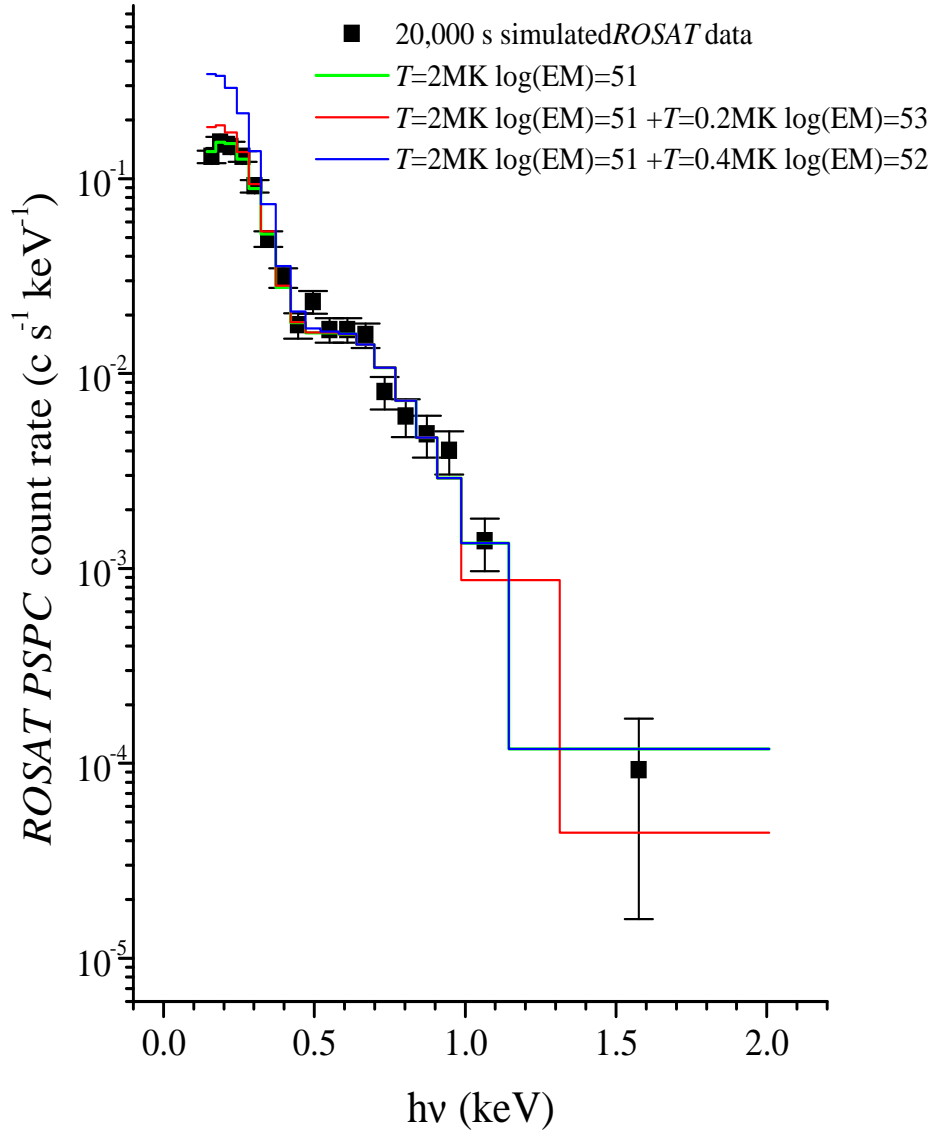}{6in}{0}{65}{70}{-200}{-40}
\caption{Three models at the resolution of the {\it ROSAT PSPC}, along with a 
simulated data set generated from the isothermal, $T = 2 \times 10^6$ K, model 
(points with error bars).  The three models include the isothermal two million 
degree K model with an emission measure of $10^{51}$ cm$^{-3}$, which is thought 
to be representative of B stars, as well as two additional models that include a 
two million degree component but also have cooler components: one with a $T = 2 
\times 10^5$ K component with 100 times the emission measure as the hotter 
component, and one with a $T = 4 \times 10^5$ K component with 10 times the 
emission measure.} 
\label{Fig:warm}
\end{figure}

If large quantities of shocked plasma do exist at the disk-wind interfaces, then 
the shocks must be relatively weak, and the temperatures correspondingly low.  
One might ask how much ``warm'' plasma ($T \approx 10^5$ K) could be present on 
Be stars and still be consistent with the X-ray data.  In Figure \ref{Fig:warm} 
I show several emission models convolved with the {\it ROSAT PSPC} spectral 
response matrix.  
It can be seen from this figure that even a huge amount of $T = 2 \times 10^5$ K 
plasma could not be reliably detected with {\it ROSAT} in the presence of the 
emission from a typical quantity of $T = 2 \times 10^6$ K plasma.  However, $T = 
4 \times 10^5$ K plasma would be detectable.  A shock temperature of 
$T=2\times10^5$ K corresponds to a shock jump velocity of 120 km s$^{-1}$, which 
is quite typical of the predictions of the oblique shocks on the surface of WCD 
disks (Bjorkman \& Cassinelli 1993). So, as Figure \ref{Fig:warm} indicates, 
large quantities of shocked gas could exist at the wind-disk interface without 
{\it ROSAT} being able to detect it.

\section{Prospects for Future Missions}

The current X-ray missions have provided a modest amount of information about 
the flux levels and time variability properties of hot stars, including Be 
stars.  However, their very limited spectral resolution has prevented any 
quantitative X-ray spectral diagnostics of Be stars from being employed.  This 
situation will change shortly with the deployment of the next generation of 
X-ray telescopes: {\it Chandra}, {\it XMM}, and {\it ASTRO-E}, which have spectral 
resolutions approaching $\lambda / \Delta \lambda \approx 1000$.  With this 
resolution, numerous individual X-ray lines will be measurable, and if they are 
wind broadened, they may even be resolvable. 

High-resolution measurements of X-ray lines (and continua) will make possible 
the application of various spectral diagnostics.  These include density- and 
temperature-sensitive line ratios, thermal versus non-thermal emission, 
deviations from equilibrium, Doppler shifts due to bulk motion, line opacity 
effects, and ionization balance via inner-shell X-ray transitions (fluorescence 
and resonant absorption).  

\begin{figure}
\centering
\epsfig{file=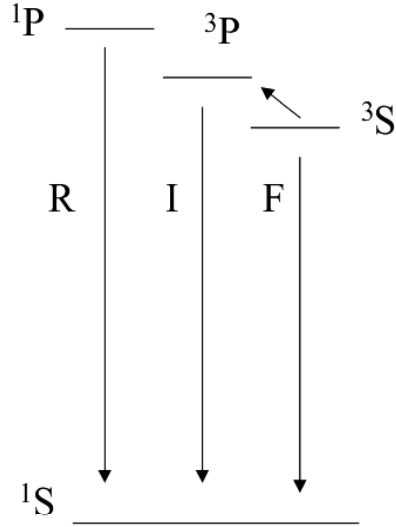, height=2.75in,clip=}
\caption{Schematic energy level diagram for a helium-like ion.  The resonance 
(R), intercombination (I), and forbidden (F) transitions are indicated.  Above 
some critical density (which increases with atomic number) collisions become 
efficient at transferring electrons from the triplet S level to the triplet P 
level, causing the intercombination line to strengthen at the expense of the 
forbidden line.} 
\label{Fig:FIR}
\end{figure}

One specific diagnostic with a lot of potential to discriminate among the 
various models for X-ray production in Be stars is the helium-like 
forbidden-to-intercombination line strength ratio, which is density sensitive.  
This physics 
behind this diagnostic is demonstrated schematically in Figure \ref{Fig:FIR}.  
This density diagnostic can be applied to helium-like ions from many elements, 
with each element having a somewhat different density- and 
temperature-sensitivity.  For example, helium-like carbon exists at 
temperatures of about $3 
\times 10^5$ K, and can diagnose densities of about $n_e = 10^9$ cm$^{-3}$, 
while helium-like silicon exists at temperatures near $3 \times 10^6$ K and its 
forbidden/intercombination line ratio is sensitive to densities near $3 \times 
10^{13}$ cm$^{-3}$.  Using these line ratios to diagnose the density of the X-
ray emitting plasma in Be stars will provide information about whether this hot 
plasma exists in the stellar wind, the disk, or even near the photosphere.  
Furthermore, density information combined with an emission measure determination 
gives information about the volume of the emitting structures.  These data 
will put some very stringent constraints on the high-energy mechanisms operating 
in Be stars. 

Another potentially useful X-ray spectral diagnostic involves probing the Be 
disks in Be/X-ray binaries with the hard X-rays from the compact companion.  
Accretion onto the compact object in these systems yields a hard X-ray continuum 
spectrum.  If the compact object can be viewed while it is behind the Be disk, 
then an absorption line spectrum will be seen.  The useful property of inner-
shell X-ray absorption is that each ionization state leads to a single 
absorption feature, with the energy of each feature monotonically increasing 
with ionization level.  Therefore, a small section of the X-ray spectrum can 
provide information about all possible ionization stages of an element such as 
silicon, or iron, in the disk.  

\section{Conclusions}

Data, primarily from {\it ROSAT}, indicates that X-ray activity may be modestly 
stronger in Be stars than in non-emission B stars, but it is qualitatively not 
that different.  There are, however, a subset of Be stars in which X-ray 
activity is quite strong and/or time variable. Taken together, these 
observational facts may indicate that the Be phenomenon has a high-energy aspect 
that can lead to enhanced X-ray emission, at least at some times and in some 
stars.  Some objects ($\lambda$ Eri, for example) have X-ray properties that 
simply cannot be understood in terms of the standard picture of wind shocks.  

The long {\it ROSAT} observation of $\beta$ Cep (and of three other $\beta$ 
Cephei stars) indicates that pulsations can have an effect on X-ray production 
in hot stars, although the specific mechanism that connects these two phenomena 
is open for debate.  As of now, however, no Be stars (save $\beta$ Cep) have 
X-ray data with sufficient time coverage to draw any conclusions about the 
modulation of X-rays by pulsations. 

The next generation of X-ray telescopes has the potential to vastly increase our 
understanding of the high-energy physical processes that occur on Be stars.  
This can be accomplished through the application of spectral diagnostics that 
are, for the most part, standard techniques in the UV, optical, and IR, but 
have not been applied in the X-ray due to the low spectral resolution of current 
instruments.

\acknowledgments

I am grateful to Joe Cassinelli, Joe MacFarlane, and Stan Owocki for providing 
useful advice and suggestions during the preparation of this paper.  This work 
was partially supported by NASA grant NAG 5-3530 to the Bartol Research 
Institute at the University of Delaware.

\section*{Discussion} 

\noindent {\bf R. Ignace:}
Are the somewhat higher $L_X$ values of Be stars over B stars consistent with Be 
winds having slightly larger mass-loss rates, as you alluded to ({\it i.e.} the 
X-rays are primarily from the polar wind, and do not arise in the disk)?

\noindent {\bf D. Cohen:}
Yes, this would be the most conservative explanation for the majority of Be 
stars.

\noindent {\bf H. Henrichs:}
Did all pulsating B stars in your sample show variable X-ray emission?

\noindent {\bf D. Cohen:}
Yes, $\beta$ Cru and $\xi^1$ CMa at $> 3 \sigma$, and $\beta$ Cep and $\alpha$ 
Lup at just below the $3 \sigma$ level.

\noindent {\bf C. Aerts:}
You observe X-ray variability in the hot $\beta$ Cephei stars, but not in 
$\alpha$ Lup, which is cooler.  You should check if this is a general 
conclusion.  Also, $\beta$ Cru is an eccentric binary in which the pulsation 
behavior seems to change between apastron and periastron.  So it would be 
interesting to know if the X-ray luminosity is different at these two times.

\noindent {\bf D. Cohen:}
Well, $\alpha$ Lup is marginally variable in our {\it ROSAT} data.  Also, I 
agree that it would be interesting to re-examine $\beta$ Cru at different 
orbital phases.   It is an interesting object, from an X-ray point of view.

\noindent {\bf A. Tarasov:}
X-ray photometry can be useful for finding hidden hot subdwarfs in these 
systems.  In this case you can see some extra emission even in cool (B2 - B5) 
stars.

\noindent {\bf D. Cohen:}
Yes, but with the very poor spectral resolution of {\it ROSAT}, and the 
inherently soft X-ray emission of the B/Be stars themselves, this is a practical 
impossibility.  But, with future high spectral resolution missions, it should 
certainly be possible to detect subdwarf companions.

\noindent {\bf M. Smith:}
You mentioned that for the possibly two good cases for X-ray variability 
($\lambda$ Eri and possibly $\mu$ Cen), it may be that these two stars are also 
unusual in having slightly more optical line 
profile variability than most other Be stars.  
So, I may be going slightly out on a limb in saying by conjecturing that the 
higher activities in these stars in both the optical and X-ray regions {\it 
might} be related.  Also, note that the ``flare'' temperature of the 1991 X-ray 
event in $\lambda$ Eri was several times hotter as well as being several times 
brighter.  This may indeed be a qualitatively different characteristic which 
would add to the challenge of modeling it with even a non-standard wind-shock 
model.

\noindent {\bf D. Cohen:}
Regarding your first point, yes, simultaneous X-ray and optical (and ideally UV) 
observations would be very nice.  On the second point, I would agree, but also 
point out that something that produces a transient increase in the X-ray flux by 
almost an order of magnitude is likely to also have a different spectral 
signature than the basal X-ray emission.  For example, the infall of a dense 
clump interacting with the outflowing stellar wind would increase the X-ray flux 
levels, but it would likely also produce
a higher X-ray temperature than the standard wind shocks.

\end{document}